\documentclass[preprint,longbibliography]{revtex4-2}
\usepackage{amsmath,amssymb,bm}
\usepackage{array}
\usepackage{graphicx}
\usepackage{enumerate}
\usepackage{color}

\usepackage[bitstream-charter]{mathdesign}
\usepackage[breaklinks=true]{hyperref}

\newcommand{\D}{\mathcal{D}}
\newcommand{\Cppk}{C_{\pi\pi}(\bm{k})}
\newcommand{\Cpps}{C_{\pi\pi}^s(\bm{k})}
\newcommand{\Cppe}{C_{\pi\pi}^e(\bm{k})}
\newcommand{\Cppr}{C_{\pi\pi}^r(\bm{k})}
\newcommand{\Cssk}{C_{\sigma\sigma}(\bm{k})}

\newcommand{\phiR}{\bar{\phi}_R}

\begin{document}
\title{
Spontaneous symmetry breaking in two dimensions under nonequilibrium laminar flows
}
\date{\today}
\author{Yuki Minami}
\affiliation{
Faculty of Engineering, Gifu University, Yanagido, Gifu 501-1193, Japan
}

\author{Hiroyoshi Nakano}
\affiliation{
Institute for Solid State Physics, University of Tokyo, Kashiwa, Chiba 277-8581, Japan
}

\begin{abstract}
We study the long-range order in two dimensions where an order parameter is advected by laminar flows such as rotational, shear, and elongational flows.
Under these flows, we analyze an ordered state of the $O(N)$ scalar model in the large-$N$ limit.
We show that the stability of the ordered state depends on the flow pattern; shear and elongational flows stabilize the long-range order but rotational flow does not.
We discuss the physical mechanism underlying our results by connecting static correlations of fluctuations and their dynamics based on the interaction representation used in quantum mechanics.
We find that advective transport induces superdiffusion under shear and elongational flows, thereby stabilizing the long-range order.
\end{abstract}

\maketitle
\newpage

\section{Introduction}
\label{sec:introduction}
The Hohenberg--Mermin--Wagner (HMW) theorem, a cornerstone of equilibrium statistical mechanics, states that the long-range order (LRO) associated with the spontaneous breaking of continuous symmetry is prohibited below two dimensions (2D)~\cite{PhysRev.158.383, PhysRevLett.17.1133}.
This theorem is applicable to various equilibrium systems, including ferromagnets~\cite{PhysRevLett.17.1133, JMathPhys.8.1064}, crystalline solids~\cite{PhysRev.176.250, PhysRevB.19.2457}, and equilibrium flocking models~\cite{PhysRevLett.125.220601}. However, proofs of the HMW theorem are valid only for systems in thermal equilibrium, leaving open the possibility of LRO under nonequilibrium conditions.
Indeed, several examples of nonequilibrium systems that violate the HMW theorem have been reported: the flocking model represented by the Vicsek model~\cite{Chate2020-ps, Vicsek1995-hb, Toner1995-qd, Toner1998-ah, Toner2012-nk, Toner2012-ki, Nishiguchi2017-ta, Mahault2019-ox, Iwasawa2021-qc, Codina2022-jg, Besse2022-su, Ikeda2024-jn, Chate2024-be, Jentsch2024-ml}, the sheared model~\cite{De_Gennes1976-cu, Corberi2003-nv, Minami2021-po, Nakano2021-bt, Nakano2021-sh, Giomi2022-ex, Giomi2022-ax, Krommydas2023-tk}, the multitemperature conserved model~\cite{Bassler1995-lw, Tauber2002-nb, Reichl2010-ck}, the nonreciprocal model~\cite{Dadhichi2020-ug, Loos2023-la}, the center-of-mass conserving model~\cite{Galliano2023-fq, Ikeda2023-it, Maire2024-we},   athermal chiral active particles~\cite{Kuroda2023-ns, Kuroda2024-cg}, and others~\cite{Bergersen1991-sl}.
The extension of the HMW theorem to nonequilibrium systems attracts significant interest in the field of statistical mechanics.

Extending the HMW theorem is challenging owing to the diverse nature of nonequilibrium fluctuations.
Generally, the spontaneous breaking of continuous symmetry, whether in equilibrium or nonequilibrium systems, gives rise to gapless fluctuations called the Nambu--Goldstone (NG) mode~\cite{PhysRev.122.345, goldstone1961field, PhysRev.127.965, minami2018spontaneous, hidaka2020spontaneous, hongo2021effective}.
In low-dimensional systems, the NG mode may cause infrared (IR) divergences that destroy LRO~\cite{chaikin1995principles, nishimori2010elements}.
For example, in equilibrium systems, the NG mode $\pi(\bm{k})$ behaves as $\langle \pi(\bm{k})\pi(-\bm{k})\rangle \sim 1/\bm{k}^2$ in wavenumber space. 
Then, the mean square fluctuation in real space is given by 
\begin{align}
\langle \pi(\bm{x})^2\rangle &\sim \int^{2\pi/a}_{2\pi/L} \frac{dk k^{d-1}}{(2\pi)^d} \frac{1}{k^2},  \label{eq:IR divergence}
\end{align}
where $d$ is the space dimension, $L$ the system size, and $a$ the microscopic grid width. 
For $d\leq 2$, the fluctuation diverges in the thermodynamic limit $L \to \infty$, thereby destroying the LRO. 
Therefore, understanding the singularity of the NG mode in $\bm{k} \to \bm{0}$ is crucial for examining the applicability of the HMW theorem in nonequilibrium systems.
However, nonequilibrium fluctuations exhibit much more anomalous and complex behaviors than equilibrium ones, depending on the specific nature of the systems considered.
Examples include   generic long-range correlations~\cite{Garrido1990-dy, Dorfman1994-cl, De_Zarate2006-xw, Nakano2022-kv, Nakano2024-th}, the hyperuniformity of density fluctuations~\cite{Torquato2018-ft, Lei2019-rl},   giant number fluctuations~\cite{Chate2006-ey, Narayan2007-ia, Zhang2010-ir, Nishiguchi2017-ta}, the anomalous suppression of critical fluctuations~\cite{Onuki1979-jt, Cates1989-io, Katz1984-eg, Leung1991-jk, Praestgaard2000-bj} and,   NG mode splitting into an infinite number of modes~\cite{Minami2021-po}.
These anomalous fluctuations may change the behaviors of the NG mode, complicating the extension of the HMW theorem.

One class of systems that violates the HMW theorem is characterized by nonequilibrium advection. The Vicsek model and its extensions fall into this category~\cite{Chate2020-ps, Vicsek1995-hb, Toner1995-qd, Toner1998-ah, Toner2012-nk, Toner2012-ki, Nishiguchi2017-ta, Mahault2019-ox, Iwasawa2021-qc, Codina2022-jg, Besse2022-su, Ikeda2024-jn, Chate2024-be, Jentsch2024-ml}, where self-propelled motion induces the advection.
Another notable example is the sheared model~\cite{De_Gennes1976-cu, Corberi2003-nv, Minami2021-po, Nakano2021-bt, Nakano2021-sh, Giomi2022-ex, Giomi2022-ax, Krommydas2023-tk}, where externally imposed shear flow advects an order parameter.
In both cases, analyses of minimal hydrodynamic models have established the occurrence of LRO in 2D.
For the Vicsek model and its extensions, the minimal hydrodynamic model is given by~\cite{Toner2012-ki, Besse2022-su, Ikeda2024-jn, Chate2024-be}:
\begin{align}
\frac{\partial \bm{u}(t,\bm{x})}{\partial t} + \lambda \bm{u}(t,\bm{x}) \cdot \nabla \bm{u}(t,\bm{x}) = - \frac{\delta F[\bm{u}]}{\delta \bm{u}} + \bm{f}(t,\bm{x}),
\label{eq:flocking model: basis}
\end{align}
where $\bm{u}(t,\bm{x})$ is the velocity field and $\bm{f}(t,\bm{x})$ is the Gaussian white noise   mimicking errors, and $F[\bm{u}]=(\nabla \bm{u})^2/2 - a\bm{u}^2/2 + b\bm{u}^4/4$ is the free energy promoting the formation of a uniform flocking state with $\bm{u}\neq \bm{0}$.
The term $\lambda (\bm{u} \cdot \nabla) \bm{u}$ represents the self-advection of $\bm{u}$ due to the self-propelled motion of microscopic elements.
The uniform flocking state appears whenever $\lambda > 0$, although the robustness and stability of this flocking phase remain debated~\cite{Codina2022-jg, Besse2022-su}.
Note that experimental observations have confirmed the existence of the two-dimensional flocking state~\cite{Nishiguchi2017-ta, Iwasawa2021-qc}.
For the sheared model, the minimal hydrodynamic model is given by~\cite{De_Gennes1976-cu, Corberi2003-nv, Minami2021-po, Nakano2021-bt, Nakano2021-sh, Giomi2022-ex, Giomi2022-ax, Krommydas2023-tk} 
\begin{align}
\frac{\partial \bm{\phi}(t,\bm{x})}{\partial t} + \bm{v}(\bm{x}) \cdot \nabla \bm{\phi}(t,\bm{x}) = - \frac{\delta F[\bm{\phi}]}{\delta \bm{\phi}} + \bm{\eta}(t,\bm{x}),  
\label{eq:sheared model: basis}
\end{align}
where $\bm{v}(\bm{x})= (\dot{\gamma}y,0)$ is the stationary shear flow,  $\bm{\phi}(t,\bm{x})$ is the $O(N)$ scalar field, $\bm{\eta}(t,\bm{x})$ is the thermal noise, and $F[\bm{\phi}]=(\nabla \bm{\phi})^2/2 - a\bm{\phi}^2/2 + b\bm{\phi}^4/4$ is the free energy promoting the formation of a spin-aligned state with $\bm{\phi}\neq \bm{0}$.
The term $\bm{v}(\bm{x}) \cdot \nabla \bm{\phi}(t,\bm{x})$ represents the advective transport.
Recent studies have shown that this minimal model describes 
$p$-atic liquid crystals~\cite{Giomi2022-ex, Giomi2022-ax}. Furthermore, theoretical analyses and numerical simulations for $N=2$ have demonstrated the stability of the LRO in 2D~\cite{Minami2021-po, Nakano2021-bt, Giomi2022-ex, Giomi2022-ax}.

\begin{figure}[t]
  \begin{center}
        \begin{center}
          \includegraphics[scale=0.8]{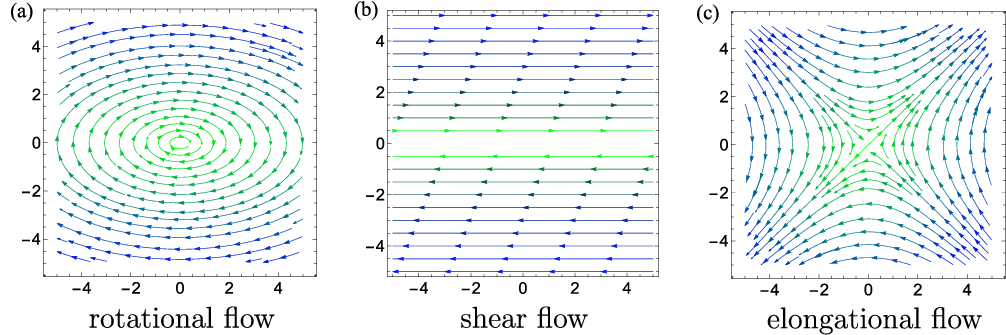}
        \end{center}
    \caption{Three flow patterns of the linear laminar flow (\ref{eq: flow}). (a) Rotational flow at $S=0.5$ and $A=1.0$. (b) Shear flow at $S=A=1.0$. (c) Elongational flow at $S=1.0$ and $A=0.0$.}
    \label{fig_general_laminar_flow}
  \end{center}
\end{figure}

Despite the clear evidence that the nonequilibrium advection of the form $\bm{u} \cdot \nabla \bm{u}$ or $\bm{v} \cdot \nabla \bm{\phi}$ can qualitatively change the dynamic behaviors and lead to the LRO in 2D, analyses using scaling theory, renormalization group theory, or numerical simulations directly solving  (\ref{eq:flocking model: basis}) or (\ref{eq:sheared model: basis}) do not fully reveal how the nonequilibrium advection plays a crucial role in the dynamics of these systems and violates the HMW theorem.

The aim of this study is to elucidate the physical mechanism underlying the emergence of the LRO in 2D due to advective transport, with the ultimate goal of extending the HMW theorem to nonequilibrium systems.
For this purpose, we focus on the externally imposed flow and generalize the shear flow 
to the laminar flow expressed as~\cite{OK1, OK2}
\begin{align}
\bm{v}(\bm{x})=\mathcal{D} \cdot \bm{x}
\quad \mbox{with} \quad
\mathcal{D}=
\begin{pmatrix}
0 & S+A \\
S-A & 0
\end{pmatrix}.
\label{eq: flow}
\end{align} 
We call   flow (\ref{eq: flow}) the linear laminar flow.
Here, $\mathcal{D}$ represents the coefficient matrix, and $S$ and $A$ are non-negative parameters.
This stationary flow includes three patterns depending on   $S$ and $A$: rotational flow ($S<A$), shear flow ($S=A$), and elongational flow ($S>A$), the streamlines of which are shown in Fig.~\ref{fig_general_laminar_flow}.
Under these various flow patterns, we study the spontaneous symmetry breaking of the $O(N)$ model.
We note that   phase transitions without symmetry breaking under the linear laminar flow, such as a liquid--gas transition, were previously studied~\cite{Onuki1979-jt, OK1, OK2}.

We demonstrate that in the large-$N$ limit, the shear and elongational flows stabilize the LRO in 2D, whereas the rotational flow does not, which are summarized in Tab.~\ref{tab:effects of flows on 2D LRO}.
This finding implies that nonequilibrium advective transport does not necessarily stabilize LRO in 2D, contrary to findings of the previous studies on the sheared and flocking models.
To understand the physical mechanism underlying our results, we connect the static correlation of the NG mode and their dynamics.
We apply the interaction representation used in quantum mechanics for decomposing the time evolution of the NG mode by advection and diffusion.
In the interaction representation, the advective transport provides the time evolution of the wavenumber along streamlines of the linear laminar flow.
We show that the interplay between the advective transport and the diffusive relaxation under the shear and elongational flows leads to the superdiffusion of the NG mode, which decays faster than normal diffusion and leads to the LRO in 2D.
\begin{table}[t]
\centering
\begin{tabular}{|>{\raggedright}p{4cm}||>{\raggedright}p{2cm}||>{\raggedright\arraybackslash}p{7cm}|}
\hline
\textbf{Type of advection} & \textbf{LRO in 2D} & \textbf{Remarks} \\ \hline\hline
Rotational flow & No & New \\ \hline
Shear flow & Yes & Shown in previous studies~\cite{Minami2021-po, Nakano2021-bt, Giomi2022-ex, Giomi2022-ax} \\ \hline
Elongational flow & Yes & New \\ \hline \hline
Self-advection & Yes & Not addressed in this paper\\ \hline
\end{tabular}
\caption{Effects of flows on   LRO in 2D}
\label{tab:effects of flows on 2D LRO}
\end{table}


\section{Setup}
\label{sec:model}
\subsection{Model}
We consider a $1+2$ dimensional $O(N)$ scalar model subjected to the linear laminar flow (\ref{eq: flow}).
Here, the order parameter is given by the $N$ component real scalar field $\phi^a(t,\bm{x})$ and its dynamics is described by the nonlinear Langevin equation:
\begin{align}
\frac{\partial }{\partial t}\phi^a(t,\bm{x})+ \bm{v}(\bm{x}) \cdot\nabla \phi^a(t,\bm{x}) &=-\Gamma \frac{\delta F}{\delta \phi^a}+\eta^a(t,\bm{x}), 
\label{eq:modelA}
\end{align}
where $a=1,2,...,N$, $\Gamma$ is the dissipation constant, and $\eta^a(t,\bm{x})$ is the Gaussian white noise with the temperature $T$, which obeys
\begin{align}
\langle 
\eta^a(t,\bm{x})\eta^b(t',\bm{y})\rangle=2T\Gamma\delta^{ab}\delta(t-t')\delta(\bm{x}-\bm{y}).
\label{eq:noise}
\end{align}
We choose the free energy of the standard form:
\begin{align}
F= \int d\bm{x}\biggl[ \frac{1}{2}\sum_a(\nabla \phi^a)^2+\frac{r}{2}\sum_a(\phi^a)^2+\frac{g }{4 N}\biggl(\sum_a(\phi^a)^2\biggr)^2\biggr].\label{eq:free energy}
\end{align}
When the term $\bm{v}(\bm{x}) \cdot\nabla \phi^a(t,\bm{x})$ is absent, our model reduces to the Ginzburg--Landau model~\cite{chaikin1995principles}.
This term gives the advective transport of the order parameter by the flow with the velocity $\bm{v}(\bm{x})$.
When we neglect the right-hand side of (\ref{eq:modelA}), it becomes an advection equation, which has the general solution
\begin{align}
\phi^a(t,\bm{x}) = \Phi^a\big(\bm{x}-\bm{v}(\bm{x})t\big),
\end{align}
for which we have assumed the initial condition $\phi^a(0,\bm{x}) = \Phi^a(\bm{x})$.
This solution means that the order parameter is transported with the velocity $\bm{v}(\bm{x})$.

Note that our model has the internal $O(N)$ symmetry of $\phi^a$. 
The equation of motion (\ref{eq:modelA}) and the noise correlation (\ref{eq:noise}) are covariant under the transformation
\begin{align}
\phi^a \to \sum_b {[R^n]^a}_b \phi_b, \quad \eta^a \to \sum_b {[R^n]^a}_b \eta_b,
\end{align}
where $R^n$ is the rotation matrix that mixes the internal degrees of freedom $a$ and $b$.
The subscript $n$ specifies the combinations of two components chosen from  $N$ components and runs $1$, $2$..., $N(N-1)/2$. 
We stress that $R^n$ mixes solely the internal degrees of freedom without mixing the spatial coordinates $x$ and $y$. 
Therefore, the advection term $\bm{v}(\bm{x}) \cdot\nabla \phi^a(t,\bm{x})$ does not violate the internal $O(N)$ symmetry, although it explicitly breaks the spatial rotation symmetry.

\subsection{Ordered state at zero temperature}
\label{eq:zero tempearture}
The steady state at $T=0$ is homogeneous in space and does not depend on the flow $\bm{v}(\bm{x})$ because the advection term vanishes for homogeneous solutions.
The flow effects appear in fluctuations at finite temperatures, as seen in the next section.
Then, the steady state is obtained by the minimum of the free energy (\ref{eq:free energy}).
The saddle point equation is given by
\begin{align}
\frac{\delta F}{\delta \phi^a}\biggl\vert_{\phi^a=\bar{\phi}^a} = \biggl[r+\frac{g}{N} \sum_b (\bar{\phi}^b)^2\biggr]\bar{\phi}^a = 0,
\end{align}
which has the disordered solution for $r\geq 0$, 
\begin{align}
\bar{\phi}^a=(0,0,...,0),
\end{align}
and   the ordered solution for $r<0$,
\begin{align}
\bar{\phi}^a = \biggl(\sqrt{\frac{-Nr}{g}},0,...,0\biggr).
\label{eq:ordered solution: T=0}
\end{align}
Here, we can choose the first component as the ordering direction without loss of generality. This solution is still invariant under $O(N-1)$ rotations that do not mix the first component.
Thus, the spontaneous symmetry breaking of $O(N) \to O(N-1)$ occurs in the ordered state.

\section{Fluctuations in the ordered state at finite temperature}
\label{sec:finite temperature}

\subsection{Formal expression of transition point}
At a finite temperature, thermal fluctuations may disrupt the ordered state at $T=0$.
To examine the thermal fluctuations under the flow, we consider the fluctuations $\sigma(t,\bm{x})$ and $\pi^\alpha(t,\bm{x})$ from the ordered state:
\begin{align}
\phi^a(t,\bm{x})=(\sqrt{N}\phiR+\sigma(t,\bm{x}), \pi^\alpha(t,\bm{x})), \hspace{1cm} \alpha=2,3,..., N,
\end{align}
where $\bar{\phi}_R$ is the condensation including renormalization corrections.
We note that $\pi^\alpha$ corresponds to the NG mode.
From the leading-order calculation in the $1/N$ expansion, we can obtain the complete set of equations describing the condensation and fluctuations in the stationary state at $t \to \infty$
\begin{align}
&\phiR^2 = -\frac{1}{g}\biggl(r+ g I_{\pi\pi}\biggr),  \label{eq:condensate} \\
&I_{\pi\pi} := \langle \pi (\bm{x})^2 \rangle =
\int \frac{d\bm{k}}{(2\pi)^2} \Cppk, \label{eq:def_Cpipi}\\
&\Cppk =  \frac{\Gamma T}{\Gamma \bm{k}^2-(1/2)\bm{k} \cdot  \D \cdot \nabla_{\bm{k}}}, \label{eq:picorrelation} \\
&\Cssk
= \frac{\Gamma T}{\Gamma(\bm{k}^2-2r-2gI_{\pi\pi})-(1/2)\bm{k} \cdot  \D \cdot \nabla_{\bm{k}}},\label{eq:sigmacorrelation}
\end{align}
where $\bm{k}$ is the wavenumber and $\nabla_{\bm{k}} := \partial / \partial \bm{k}$. Here, we have introduced the static correlation $C_{XX}(\bm{k})$ by
\begin{align}
   \langle X(\bm{k}) X(\bm{k}')\rangle = C_{XX} (\bm{k}) (2\pi)^2 \delta(\bm{k}+\bm{k}').
\end{align}
We abbreviate the superscript $\alpha$ of $\pi^\alpha(\bm{k})$ in $\Cppk$ because $\pi^{\alpha}$ are equivalent for all $\alpha$ owing to the $O(N-1)$ symmetry. 
The expressions~(\ref{eq:condensate})--(\ref{eq:sigmacorrelation}) are valid for any $S$ and $A$.
The deviation is given in appendix~\ref{sec:derivation1}.
Note that $I_{\pi\pi}$ gives the one-loop renormalization corrections to $\phiR^2$ and $\Cssk$.
In particular, we have the renormalized "mass" of $\sigma$ as
\begin{align}
m_\sigma^2=-2r-2gI_{\pi\pi} 
\label{eq:renormalized mass}
\end{align}
from the denominator of (\ref{eq:sigmacorrelation}) in $\bm{k}\to \bm{0}$.
In contrast, $\pi^\alpha$ is protected from   mass renormalization because it is the NG mode.
We note that the renormalization correction $I_{\pi\pi}$ vanishes at $T\to 0$, and then we recover the condensate (\ref{eq:ordered solution: T=0}) at $T=0$.

The transition point between the ordered and disordered phases is determined by the stability of the ordered state.
The ordered state is stable if the renormalized mass squared is non-negative $m_\sigma^2 \geq 0$, whereas it is unstable if the renormalized mass squared is negative $m_\sigma^2 < 0$.    
Then, from (\ref{eq:renormalized mass}), we have the transition point $r^*$ as
\begin{align}
r^* &= -g I_{\pi \pi},\label{eq:transitionpoint}\\ 
&= -g\int \frac{d\bm{k}}{(2\pi)^2} \Cppk, \label{eq:transitionpoint}
\end{align}
where we use (\ref{eq:def_Cpipi}) in the second equality.
This equation provides the relationship between the transition point $r^*$ and the static correlation $\pi (\bm{k})$ of the NG mode. 
The transition point $r^{*}$ may diverge to $-\infty$ owing to the IR singularities of $\Cppk$.
Because the ordered state is unstable for $r>r^*$, the negative divergence of $r^*$ implies that the ordered state is not realized for any finite $r$.
Consequently, it turns out that the realization of the ordered state can be identified by examining the IR divergence of the integral  (\ref{eq:transitionpoint}).

In fact, in the equilibrium case with $\mathcal{D}=0$, $\Cppk$ is calculated as $T/\bm{k}^2$ from (\ref{eq:picorrelation}) and results in the logarithmic divergence of $r^{*}$ in 2D:
\begin{align}
r^* &= -g\int^{2\pi/a}_{2\pi/L} \frac{d\bm{k}}{(2\pi)^2} \frac{T}{\bm{k}^2}, \\ 
&= -\frac{gT}{2\pi} \log \frac{L}{a} \to -\infty. 
\label{eq:IR divergence of r}
\end{align}
This result is consistent with the   HMW theorem (\ref{eq:IR divergence}).

\subsection{Interplay between advection and diffusion}
\begin{figure}[t]
  \begin{center}
        \begin{center}
          \includegraphics[scale=0.8]{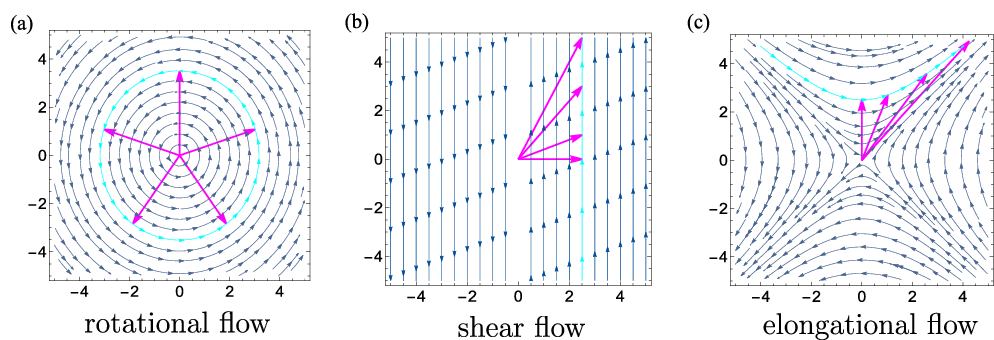}
        \end{center}
    \caption{Time evolution of the advected wavenumber $\bm{q}(t)$. The magenta arrows show $\bm{q}(t)$ at several times. The dark- and light-blue lines show the streamlines of $\tilde{\bm{v}}(\bm{k})$ and the line passing through $\bm{q}(0)$, respectively. 
    (a): $\bm{q}(t)$ under the rotational flow with $S=0.1$ and $A=1.0$.
    (b): $\bm{q}(t)$ under the shear flow with $S=A=0.2$.
    (c): $\bm{q}(t)$ under the elongational flow with $S=1$ and $A=0$. 
    }
    \label{fig:q(t)}
  \end{center}
\end{figure}
To evaluate the IR divergence of  (\ref{eq:transitionpoint}) for the various flow patterns, we derive the integral expression of $\Cppk$ from   (\ref{eq:picorrelation}).
To this end, we exponentiate the denominator of  (\ref{eq:picorrelation}) by introducing the time integral
\begin{align}
\Cppk &= \int^\infty_0 dt U(t) \Gamma T,\label{eq:Cpipi}\\
U(t)&:=\exp\biggl[t\biggl(-\Gamma\bm{k}^2+\frac{1}{2}\bm{k} \cdot  \D \cdot \nabla_{\bm{k}}\biggr)\biggr],\label{eq:U}
\end{align}
As we will see in section~\ref{sec:physical interpretation}, $U(t)$ corresponds to the time-evolution operator of the NG mode. Then, we can regard the exponent of $U(t)$ as the Hamiltonian governing the time evolution:
\begin{align}
H=-\Gamma\bm{k}^2+\frac{1}{2}\bm{k} \cdot  \D \cdot \nabla_{\bm{k}}.
\end{align}
This Hamiltonian consists of two parts:
\begin{align}
H_I=-\Gamma \bm{k}^2 \hspace{0.2cm}\textrm{and}\hspace{0.2cm} H_0=\frac{1}{2}(\D^T\cdot \bm{k}) \cdot \nabla_{\bm{k}}.
\end{align}
The first part $H_I$ describes diffusive relaxation, whereas the second part $H_0$ describes advective transport.
The interplay between $H_I$ and $H_0$ results in the rich behaviors of $U(t)$ or $\Cppk$. 
To decompose the roles played by diffusion and advection, we apply the interaction representation used in quantum mechanics and rewrite (\ref{eq:U}) as
\begin{align}
U(t)&=U_I(t)e^{tH_0},\label{eq:interaction picture}\\
U_I(t)&= \exp\biggl[-\Gamma\int^t_0 d\tau \bm{q}(\tau)^2\biggr],\label{eq:UI} \\
\bm{q}(\tau) &:= e^{\tau H_0} k e^{-\tau H_0}.\label{eq:def of q}
\end{align}
See Appendix~\ref{sec:U(t)} for details.
Then, the correlation function becomes
\begin{align}
\Cppk = \Gamma T\int^\infty_0 dt U_I(t) = \Gamma T\int^\infty_0 dt \exp\biggl[-\Gamma\int^t_0 d\tau \bm{q}(\tau)^2\biggr]
\label{eq:picorrelation_2}
\end{align}
where $e^{tH_0}$ of (\ref{eq:interaction picture}) becomes unity because it acts on the constant $\Gamma T$ in (\ref{eq:Cpipi}). 
Here, we introduce the advected wavenumber $\bm{q}(\tau)$, which evolves with time according to only the advective transport $H_0$.
In the representation (\ref{eq:picorrelation_2}), the effects of advection are fully encapsulated in $\bm{q}(t)$, whereas the quadratic form of $\bm{q}(\tau)^2$ reflects the effects of diffusion.

\subsection{Stability of the ordered state}
\label{sec:stab}
From (\ref{eq:transitionpoint}) and (\ref{eq:picorrelation_2}), we calculate the long-wavelength behaviors of $\Cppk$ and the stability of the ordered state at each flow.
We first note that from (\ref{eq:def of q}), $\bm{q}(t)$ is the solution of the advection equation:
\begin{align}
\frac{\partial}{\partial t} \bm{q}(t) &= \tilde{\bm{v}}(\bm{k})\cdot \nabla_k \bm{q}(t), \label{eq:advection equation:1}
\\
\tilde{\bm{v}}(\bm{k})&=\frac{1}{2}\D^T \cdot \bm{k},
\label{eq:advection equation:2}
\end{align}
with the initial condition $\bm{q}(0) = \bm{k}$.
Then, $\bm{q}(t)$ evolves along the streamlines of $\tilde{\bm{v}}(\bm{k})$ without the diffusion. 
In Fig.~\ref{fig:q(t)}, we plot the time evolution of $\bm{q}(t)$, which significantly depends on the flow pattern.
In the long-time limit, we can calculate the asymptotic behaviors of $\bm{q}^2(t)$ as 
\begin{align}
    \bm{q}^2_s(t) \sim t^2 k_x^2, \quad \bm{q}^2_e(t) \sim e^{t\Omega_e} k_{\rm out}^2, \quad 
    \bm{q}^2_r(t) \sim k^2. 
\label{eq:long-time behavior of q(t)}
\end{align}
where the subscripts $s$, $e$, and $r$ denote the shear, elongational, and rotational flows, respectively.
Here, $\Omega_e=\sqrt{S^2-A^2}$ and $k_{\rm out}$ is the wavenumber along the outgoing direction of the elongational flow in Fig.~\ref{fig:q(t)}-(b).
See appendix~\ref{sec:q(t)} for   details.
Then, substituting (\ref{eq:long-time behavior of q(t)}) into (\ref{eq:UI}) yields the dependence of $U_I(t)$ on the flow patterns 
\begin{align}
U_I^s(t) \sim e^{-\Gamma S^2 k_x^2 t^3}, \quad U_I^e(t) \sim e^{-\Gamma S^2\Omega_e^{-1}k_{\rm out}^2 e^{ t\Omega_e }}, \quad U_I^r(t) \sim e^{-\Gamma\bm{k}^2 t}. \label{eq:UI in long time}
\end{align}
By integrating (\ref{eq:UI in long time}) in time, we can derive the asymptotic expressions of $\Cppk$ from (\ref{eq:picorrelation_2}). 
We first consider the time integrals of $U_I^s(t)$ and $U_I^r(t)$.
We extract the wavenumber dependences of the time integrals of $U_I^s(t)$ and $U_I^r(t)$ by changing the time variables as
\begin{align}
    \int^\infty_0 dt U_I^s(t) &\sim \int^\infty_0 dt e^{-\Gamma S^2 k_x^2 t^3} = \frac{1}{k_x^{2/3}} \int^\infty_0 d\tau_s e^{-\Gamma S^2 \tau_s^3}, \label{eq:calc of UIs}\\
     \int^\infty_0 dt U_I^r(t) &\sim \int^\infty_0 dt e^{-\Gamma k^2 t} = \frac{1}{k^2} \int^\infty_0 d\tau_r e^{-\Gamma \tau_r}, \label{eq:calc of UIr}
\end{align}
where $\tau_s=k_x^{2/3} t$ and $\tau_r = k^2 t$.
We note that the integral parts of the right-hand side do not depend on the wavenumber.
Thus, we have the asymptotic behaviors
\begin{align}
\Cpps &\sim \frac{1}{k_x^{2/3}}, \label{eq:Cpps asym}\\
\Cppr &\sim \frac{1}{\bm{k}^2}.\label{eq:Cppr asym}
\end{align}
We next consider the time integral of $U_I^e(t)$.
By introducing the time variable as
\begin{align}
  \tau_e=\Gamma S^2 \Omega_e^{-1}k_{\rm out}^2 e^{t\Omega_e},
\end{align}
we have
\begin{align}
    \int^\infty_0 dt U_I^e(t) \sim \int^\infty_0 dt \exp\biggl[-\Gamma S^2 \Omega_e^{-1} k^2_{\rm out} e^{t\Omega_e}\biggr]
    &=\frac{1}{\Omega_e}\int^\infty_{\Gamma S^2 \Omega_e^{-1} k^2_{\rm out}}\frac{e^{-\tau}}{\tau},\\
    &=-\frac{1}{\Omega_e}\mbox{Ei}(-\Gamma S^2 \Omega_e^{-1} k^2_{\rm out}), \label{eq:Ei}
\end{align}
where Ei$(x)$ is the exponential integral function
\begin{align}
    \mbox{Ei}(x)=-\int^\infty_{-x}d\tau \frac{e^{-\tau}}{\tau}.
\end{align}
The exponential integral function is represented as~\cite{abramowitz1968handbook}
\begin{align}
\mbox{Ei}(x)=\log x+\gamma-\mbox{Ein}(-x),
\end{align}
where $\gamma$ is the Euler constant and Ein$(x)$ is the entire function. 
Because Ein$(x)$ is a regular function, the singularity of (\ref{eq:Ei}) in the limit $k_{\rm out} \to 0$ is given by
\begin{align}
      \int^\infty_0 dt U_I^e(t)&\sim \int^\infty_0 dt \exp\biggl[-\Gamma S^2 \Omega_e^{-1} k^2_{\rm out} e^{t\Omega_e}\biggr],\\
    &\sim -\frac{1}{\Omega_e}\log k_{\rm out}^2. \label{eq:Cpps asym}
\end{align}
The resulting asymptotic forms of $\Cppk$ are summarized as
\begin{align}
\Cpps \sim \frac{1}{k_x^{2/3}}, \quad \Cppe \sim -\log|k_{\rm out}|, \quad \Cppr \sim \frac{1}{\bm{k}^2}.
\label{eq:static NG mode: list}
\end{align}

The rotational flow does not change the asymptotic form of $\Cppk$ from that of equilibrium.
Then, as shown in (\ref{eq:IR divergence of r}), the transition point $r^{*}$ diverges to $-\infty$ and the ordered state does not appear for any finite $r$.
We note that the rotational flow includes both purely rotational ($S=0$) and elliptic ($S\neq 0$) flows. 
The elliptic flow drives the systems out of equilibrium and explicitly breaks a spatial rotational symmetry. Then, our result shows that the nonequilibrium conditions do not necessarily stabilize the LRO in 2D.
In contrast, for the purely rotational flow, the absence of the ordered state is trivial because the system under this flow is equivalent to the equilibrium systems observed from rotating frames.

The shear and elongational flows significantly change the asymptotic behaviors of $\Cppk$.
The divergent behaviors of $1/k_x^{2/3}$ and $-\log|k_{\mathrm{out}}|$ are weaker than the equilibrium behavior $1/\bm{k}^2$, implying that the fluctuations are anomalously suppressed by the flows.
We recall that as shown in (\ref{eq:IR divergence of r}), the IR divergence due to the $1/\bm{k}^2$ behavior is marginal in 2D; the weaker divergences under the shear and elongational flows do not lead to the IR divergence of $r^{*}$.
As a result, the LRO in 2D can stably exist for finite $r$.
Therefore, the HWM theorem is violated under the shear and elongational flows.
The results are summarized in Tab.~\ref{tab:effects of flows on 2D LRO}.

\section{Phase diagram}
\label{Phase diagram}
\begin{figure}[tb]
  \begin{center}
        \begin{center}
          \includegraphics[scale=0.82]{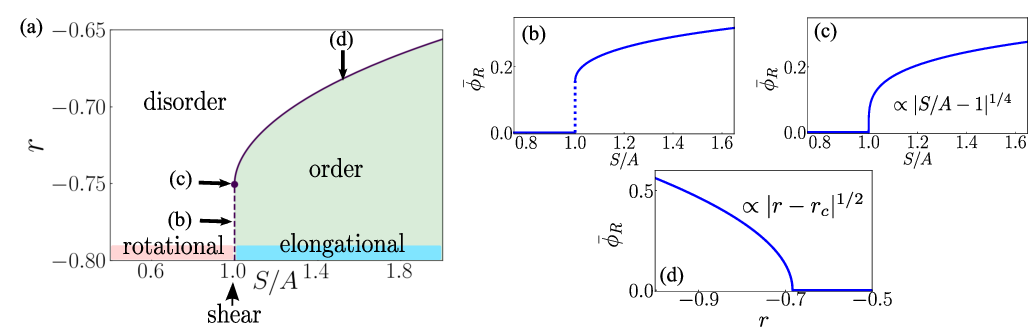}
        \end{center}
    \caption{(a): Phase diagram in $(r,S/A)$ space. (b)--(d): $S/A$ or $r$ dependence of the order parameter around the phase boundaries that are specified in (a). In (a), the phase boundary indicated by the dotted line is of the first order, whereas that indicated by the solid curve is of the second order. In all figures, we set the parameters to $g=\Gamma=T=A=1$.}
    \label{fig:phasediagram}
  \end{center}
\end{figure}
We numerically calculate the transition point and draw the phase diagram.
From   (\ref{eq:transitionpoint}) and (\ref{eq:picorrelation_2}), we obtain an explicit expression of the transition point
\begin{align}
r^* = -gT \int \frac{d\bm{k}}{(2\pi)^2}\int^\infty_0 dt \exp\biggl[-\Gamma\int^t_0 d\tau \bm{q}(\tau)^2\biggr]. \label{eq:transitionpoint: for numerical}
\end{align}
We perform   numerical integration to obtain the phase boundary $r^*(S, A)$ as a function of $S$ and $A$. See appendix~\ref{appendix:phase diagram} for the detailed procedure.
In Fig.~\ref{fig:phasediagram} (a), we present the phase diagram in $(r,S/A)$ space with $A=1$.
We note that the flow patterns can be controlled by changing the value of $S/A$. 
The phase transition occurs in the parameter range of the shear  ($S/A=1$) and  elongational  ($S/A>1$) flows, whereas in the regime of rotational ($S/A<1$) flow, the transition point $r^*$ diverges to $-\infty$ and no ordered phase appears.

In Fig.~\ref{fig:phasediagram} (b)--(d), we plot the onset behavior of the order parameter for different combinations of $r$ and $S/A$. Three characteristic types of behavior are observed in these parameter values.
The first type presented in Fig.~\ref{fig:phasediagram} (b) occurs when the flow pattern changes from rotational ($S/A<1$) to elongational ($S/A>1$) under a sufficiently small $r$ [dotted line in Fig.~\ref{fig:phasediagram} (a)].
In this case, the order parameter exhibits a discontinuous jump at $S/A=1$, indicating that the phase transition is of the first order. 
As $r$ increases, the jump width approaches $0$ and a continuous phase transition appears at the endpoint [(c) in Fig.~\ref{fig:phasediagram} (a)], which is the second type presented in Fig.~\ref{fig:phasediagram} (c).
The onset behavior at the endpoint is described by
\begin{align}
\phiR \propto \Biggl(\frac{S}{A}-1\Biggr)^{1/4}.\label{eq:exponent1}
\end{align}
Finally, the third type presented in Fig.~\ref{fig:phasediagram} (d) occurs at the phase boundary where $r$ decreases with the shear and elongational flows kept constant [solid curve in Fig.~\ref{fig:phasediagram} (a)]. In this case, the order parameter behaves as
\begin{align}
\phiR \propto |r-r^*|^{1/2}.\label{eq:exponent2}
\end{align}
These critical exponents coincide with the mean-field values because the IR divergence is suppressed.
See appendix~\ref{sec:scaling} for the derivation of the exponents.

\section{Physical mechanism underlying the violation of the HWM theorem}
\label{sec:physical interpretation}
We discuss the physical mechanism underlying our results on the basis of (\ref{eq:Cpipi}), which connects the static correlation of the NG mode to their dynamics.
We show that the shear and elongational flows cause the superdiffusion of the NG mode, which decays faster than normal diffusion and leads to the LRO in 2D.

The time correlation function of the NG mode obeys the following equation of motion
\begin{align}
\biggl[\frac{\partial}{\partial t}-\frac{1}{2}\bm{k} \cdot \D\cdot \nabla_{\bm{k}}+\Gamma \bm{k}^2\biggr]D_{\pi\pi}(t,\bm{k})=0,
\end{align}
where $D_{\pi\pi}(t,\bm{k}):=\langle \pi^\beta(t,\bm{k})\pi^\beta(0,-\bm{k})\rangle/L^2$.
The derivation is given in appendix~\ref{sec:derivation1}.
We can write the formal solution using the time evolution operator $U(t)$ of  (\ref{eq:U}) as
\begin{align}
D_{\pi\pi}(t,\bm{k}) &= U(t) D_{\pi\pi}(0,\bm{k}).  \label{eq:decay of pi}
\end{align}
Therefore, (\ref{eq:Cpipi}) is interpreted as the formula that connects the static correlation of the NG mode with their dynamics.
Because (\ref{eq:Cpipi}) remains valid even in equilibrium ($\bm{v}=\bm{0}$), it allows us to study how the nonequilibrium dynamics changes the behavior of the static correlation of the NG mode. 

\begin{table}[b]
\centering
\begin{tabular}{|>{\raggedright}p{3cm}||>{\raggedright}p{3.5cm}|>{\raggedright}p{4cm}|>{\raggedright\arraybackslash}p{2.5cm}|}
\hline
\textbf{Type of advection} & \textbf{Streamline} & \textbf{Dynamics of NG mode} & \textbf{ LRO in 2D} \\ \hline\hline
Rotational & closed curve & normal diffusion & No \\ \hline
Shear& infinitely outward & superdiffusion & Yes \\ \hline
Elongational & infinitely outward & superdiffusion & Yes\\ \hline
\end{tabular}
\caption{Relationship between dynamics of NG mode and  LRO in 2D}
\label{tab:relationship between dynamics of NG mode and occurrence of 2D LRO}
\end{table}

We then examine the diffusion process of the NG mode described by $U(t)$.
By rewriting (\ref{eq:decay of pi}) using (\ref{eq:interaction picture}), we obtain
\begin{align}
D_{\pi\pi}(t,\bm{k}) &= U_I(t) e^{tH_0} D_{\pi\pi}(0,\bm{k}), \\
&= \exp\biggl[-\Gamma\int^t_0 d\tau \bm{q}(\tau)^2\biggr] D_{\pi\pi}(t=0,\bm{q}(t)). \label{eq:decay of pi: rewrite}
\end{align}
This equation indicates that $D_{\pi\pi}(t,\bm{k})$ exhibits the diffusion governed by $U_I(t)$, whereas $e^{tH_0}$ only advects the wavenumber of the initial state.

We have obtained the long-time behaviors of $U_I(t)$ in (\ref{eq:UI in long time}).
For the rotational flow, the time evolution operator behaves as $U_I^r(t) \sim e^{-\Gamma\bm{k}^2 t}$, implying that the NG mode experiences the normal diffusion as in equilibrium.
In contrast, for the shear and elongational flows, the evolution operators behave as $U_I^s(t) \sim e^{-\Gamma S^2 k_x^2 t^3}$ and $U_I^e(t) \sim e^{-\Gamma S^2\Omega_e^{-1}k_{\rm out}^2 e^{ t\Omega_e }}$, respectively. 
These decays are faster than the normal diffusion and are called   superdiffusion. 
Note that $U_I^s(t) \sim e^{-\Gamma S^2 k_x^2 t^3}$ corresponds to the sublinear dispersion $\omega \sim k_x^{2/3}$ of the NG mode under the shear flow~\cite{Minami2021-po}.

Recall that the normal diffusion results in the marginal divergence of the transition point, as shown in (\ref{eq:IR divergence of r}) and (\ref{eq:calc of UIr}). 
Superdiffusion, which decays faster than the normal diffusion, suppresses the marginal divergence of the fluctuations and stabilizes the LRO in 2D.
We have shown in the latter part of section~\ref{sec:stab} that the superdiffusion results in finite $r^*$.  
This is the physical mechanism underlying the violation of the HWM theorem in our model.
The results are summarized in Tab.~\ref{tab:relationship between dynamics of NG mode and occurrence of 2D LRO}

Note that whether   superdiffusion occurs can be determined solely by examining the geometry of the streamlines associated with each type of flow.
In the rotational flow [Fig.~\ref{fig:q(t)}(a)], the streamlines form closed curves, whereas in the shear and elongational flows [Figs.~\ref{fig:q(t)}(b) and (c)], they form open curves extending towards infinity.
The occurrence of   superdiffusion is crucially dependent on the streamline being open.
See appendix~\ref{sec:superdiff and geometry} for a detailed discussion.

\section{Concluding remarks}
\label{sec:concluding remarks}
We have studied the $1+2$ dimensional $O(N)$ scalar model under different types of flow: shear, elongational, and rotational. 
We have demonstrated that the shear and elongational flows stabilize the LRO in 2D that is absent in   equilibrium systems, according to the HMW theorem.
The significant point of our analysis is to explicitly establish the connection (\ref{eq:Cpipi}) between the static correlations of the NG mode and their dynamics based on an interaction representation.
In the interaction representation, the advective effect is encapsulated in the advected wavenumber $\bm{q}(t)$, which evolves in time along the streamline of flows as shown in Fig.~\ref{fig:q(t)}. 
We have shown that the interplay between the advection and diffusion causes   superdiffusion and thereby stabilizes the LRO in 2D under the shear and elongational flows.
In contrast, the rotational flow gives the normal diffusion as in equilibrium and does not stabilize the LRO in 2D.
We have also numerically calculated the transition point and drawn the phase diagram.
We have found that the order of the phase transition depends on   $r$, as shown in Fig.~\ref{fig:phasediagram}.


Recently, it has been reported that in  flocking hydrodynamics, the nonlinear interactions induced by the advective transport play a crucial role in stabilizing the LRO in 2D~\cite{Toner1995-qd, Toner1998-ah, Toner2012-nk, Toner2012-ki, Ikeda2024-jn, Chate2024-be, Jentsch2024-ml}. 
In the flocking hydrodynamics, the advection arises from the self-propelled motions of active matter.
This is different from our model, where the advection is induced by the external flow.
However, both models share the common feature that the advective transport gives rise to the LRO in 2D that is absent in   equilibrium systems.
Therefore, it is interesting to establish the relationship between the self-advection in the active matter and the flow advection in our models.
We leave it as our future work.

We note that fluctuations generally tend to be suppressed in the large-$N$ limit and become larger at finite $N$.
In some models such as the $SU(N)$ Thirring model, the NG mode is not included in the large-$N$ limit and we have to include it manually for a correct analysis~\cite{witten1978chiral}.
However, in the $O(N)$ model, (\ref{eq:condensate}) and (\ref{eq:sigmacorrelation}) have the one-loop corrections of the NG mode even in the large-$N$ limit.
Then, we expect that the ordered state of our model will not be broken even at finite $N$.
Indeed, the LRO under the shear flow at $N=2$ has been established by the sophisticated numerical simulation~\cite{Nakano2021-bt}.
In addition, the superdiffusion of the elongational flow is much faster than the shear flow, whereas the rotational flow cannot stabilize even in the large-$N$ limit. 
Therefore, our results will be qualitatively valid at finite $N$.


We finally remark on the possible realization of our results in experiments.
We expect our setup to be implemented in the synchronization transition of chemical oscillation reactions~\cite{kuramoto2003chemical, toth2006collective}.
In the uniform oscillation of the synchronization transition, the $U(1)$ phase symmetry is spontaneously broken and it is possible to induce steady flows in a chemical reaction solution.
In particular, the two-dimensional elongational flow has recently been  realized in an experiment~\cite{elonge}.
Thus, from the results presented in this paper, we expect that the uniform oscillation under the rotational flow will be disturbed as the system size increases, whereas it is sustained under the shear and   elongational flows.

\section*{Acknowledgements}
We are grateful to Keiji Saito and Yoshimasa Hidaka for valuable discussions.
This work is supported by JSPS KAKENHI (Grant Numbers JP19H05791, JP21J00034, and JP22K13978).
We also thank the Yukawa Institute for Theoretical Physics at Kyoto University and RIKEN iTHEMS. Discussions during the workshop (YITP-T-24-04) on “Advances in Fluctuating Hydrodynamics: Bridging the Micro and Macro Scales” were useful to complete this work.

\begin{appendix}

\section{Correlation functions in   \texorpdfstring{$1/N$}{TEXT} expansion}
\label{sec:derivation1}

We derive   (\ref{eq:condensate}), (\ref{eq:picorrelation}), and (\ref{eq:sigmacorrelation}).
For later convenience, we recast the equation of motion  (\ref{eq:modelA}) into the form
\begin{align}
\frac{\partial }{\partial t}\phi^a&=-\Gamma \biggl[\chi_0^{-1}+\frac{g}{N}\sum_b(\phi^b)^2 \biggr] \phi^a+\eta^a,  \label{EOM}\\
\chi_0^{-1}&=\bm{x} \cdot \frac{\D^T}{\Gamma}\cdot \nabla-\nabla^2+r.
\end{align}
We consider the fluctuations $\sigma(t,\bm{x})$ and $\pi^\alpha(t,\bm{x})$ from $\phiR$ in the ordered state:
\begin{align}
\phi^a(t,\bm{x})=(\sqrt{N}\phiR+\sigma(t,\bm{x}), \pi^\alpha(t,\bm{x})), \label{fluctuations}
\end{align}
where $\alpha=2,3,...,N$.
Substituting  (\ref{fluctuations}) into  (\ref{EOM}), we obtain
\begin{align}
\biggl[\frac{1}{\Gamma} \frac{\partial }{\partial t}&+\chi_0^{-1}+3g\phiR^2
+\frac{g}{N}\biggl(\sum_\alpha (\pi^\alpha)^2+\sigma^2
\biggr)\biggr] \sigma \nonumber \\
&+\sqrt{N}\phiR \biggl[r+g\phiR^2+\frac{g}{N}\biggl(\sum_\alpha (\pi^\alpha)^2+3\sigma^2\biggr)\biggr]
=\frac{\eta^1}{\Gamma}, \label{EOMsigma0} \\
\biggl[\frac{1}{\Gamma} \frac{\partial }{\partial t}&+\chi_0^{-1}+g\phiR^2
+\frac{g}{N}\biggl(\sum_\alpha (\pi^\alpha)^2+\sigma^2+2\sqrt{N}\phiR\sigma
\biggr)\biggr] \pi^\beta =\frac{\eta^\beta}{\Gamma}. \label{EOMpi0}
\end{align}

We calculate the correlation functions in $1/N$ expansion, and thus, counting $N$ is important.
We note that the $\pi^\alpha$ correlation yields the factor $N-1$  as
\begin{align}
\sum_\alpha \langle (\pi^\alpha)^2\rangle =(N-1) \langle (\pi^\beta)^2 \rangle,
\label{factorN}
\end{align}
where $\pi^\beta$ is an arbitrary component and equivalent to the other components owing to the $O(N-1)$ symmetry.
In contrast, the number of $\sigma$ is one, and thus, $\langle \sigma^2\rangle$ does not yield the factor $N$.

Then, for the leading order calculation, we can replace the nonlinear term with
\begin{align}
\frac{1}{N}\biggl(\sum_\alpha (\pi^\alpha)^2 +\sigma^2\biggr)&\sim  \langle (\pi^\beta)^2 \rangle,  \\
&\equiv I_{\pi\pi},
\end{align}
and obtain
\begin{align}
\biggl[\frac{1}{\Gamma}\frac{\partial }{\partial t}+\chi_0^{-1}+3g\phiR^2
+g I_{\pi\pi} \biggr] \sigma +\sqrt{N}\phiR \biggl[r+g\biggl(\phiR^2+I_{\pi\pi}\biggr)\biggr]
&=\frac{\eta^1}{\Gamma}, \label{EOMsigma} \\
\biggl[\frac{1}{\Gamma}\frac{\partial }{\partial t}+\chi_0^{-1}+g\biggl(\phiR^2
+I_{\pi\pi}\biggr)\biggr] \pi^\beta &=\frac{\eta^\beta}{\Gamma}, \label{EOMpi}
\end{align}
where we have discarded the subleading terms in $1/N$. 

We first derive  (\ref{eq:condensate}) for $\phiR$.
By taking the noise average of  (\ref{EOMsigma}), we obtain
\begin{align}
\phiR\biggl[r+g\biggl(\phiR^2
+I_{\pi\pi}\biggr)\biggr]=0.
\end{align}
Therefore, we obtain  (\ref{eq:condensate}) as $\phiR \neq 0$.
In addition, by using (\ref{eq:condensate}),   (\ref{EOMsigma}) and (\ref{EOMpi}) respectively become
\begin{align}
\biggl[\frac{\partial }{\partial t}+\bm{x} \cdot \D\cdot \nabla-\Gamma \biggl(\nabla^2-2r-2gI_{\pi\pi}\biggr) \biggr] \sigma(t,\bm{x})
&=\eta^1 (t,\bm{x}), \label{EOMsigma2} \\
\biggl[\frac{\partial }{\partial t}+\bm{x} \cdot \D\cdot \nabla-\Gamma \nabla^2\biggr] \pi^\beta(t,\bm{x}) &=\eta^\beta(t,\bm{x}). \label{EOMpi2}
\end{align}

We perform the Fourier transform on the space
\begin{align}
\biggl[\frac{\partial }{\partial t}-\bm{k} \cdot \D\cdot \nabla_{\bm{k}}+\Gamma \biggl(\bm{k}^2+2r+2gI_{\pi\pi}\biggr) \biggr] \sigma(t,\bm{k})
&=\eta^1 (t,\bm{k}), \label{EOMsigma3} \\
\biggl[\frac{\partial }{\partial t}-\bm{k} \cdot \D\cdot \nabla_{\bm{k}}+\Gamma \bm{k}^2\biggr] \pi^\beta(t,\bm{k}) &=\eta^\beta(t,\bm{k}). \label{EOMpi3}
\end{align}
By multiplying  (\ref{EOMsigma3}) by $\sigma(t,\bm{k}')$ and taking the noise average, we obtain
\begin{align}
\biggl[\frac{1}{2}\frac{\partial }{\partial t}-\frac{1}{2}\bm{k} \cdot \D\cdot \nabla_{\bm{k}}+\Gamma \biggl(\bm{k}^2+2r+2gI_{\pi\pi}\biggr) \biggr] C_{\sigma\sigma}(\bm{k},t) 
&=T, 
\end{align}
where we have introduced $C_{\sigma\sigma}(\bm{k},t)$ in $\langle \sigma(\bm{k},t) \sigma(\bm{k}',t) \rangle=C_{\sigma\sigma}(\bm{k},t) \delta(\bm{k}+\bm{k}')$.
We have also used $\langle \eta^1(t,\bm{k})\sigma(t,-\bm{k}) \rangle =T$.
We can drop the time derivative term for the steady state at $t \to \infty$ and obtain
\begin{align}
\Cssk
&=\frac{T}{-(1/2)\bm{k} \cdot \D\cdot \nabla_{\bm{k}}+\Gamma \biggl(\bm{k}^2+2r+2g\biggr)}. \label{EOMsigmacorr}
\end{align}
We also obtain  $\Cppk$ (\ref{eq:picorrelation}) by repeating the similar calculation for (\ref{EOMpi3}).

Finally, we consider the time-correlation function  $D_{\pi\pi}(t,\bm{k})$ defined by 
\begin{align}
D_{\pi\pi}(t,\bm{k}) := \frac{1}{L^2}\langle \pi^\beta(t,\bm{k})\pi^\beta(0,-\bm{k})\rangle. 
\end{align}
By multiplying  (\ref{EOMpi3}) by $\pi^\beta(0,-\bm{k})$ and taking the noise average, we obtain
\begin{align}
\biggl[\frac{\partial }{\partial t}-\frac{1}{2}\bm{k} \cdot \D\cdot \nabla_{\bm{k}}+\Gamma \bm{k}^2\biggr]D_{\pi\pi}(t,\bm{k})  &=0.
\end{align}

\section{Time-evolution operator and interaction representation}
\label{sec:U(t)}

We consider the time-evolution operator $U(t)$ with the Hamiltonians $H_0$ and $H_I$:
\begin{align}
    U(t)
    &= \exp\biggl[tH_0+tH_I\biggr],\\
    H_0 &=\bm{k}\cdot \mathcal{D} \cdot \nabla_k/2,\\
    H_I &=-\Gamma \bm{k}^2.
\end{align}
To discuss  the role of  advection, we decompose the time evolution into those by the diffusion $H_0$ and the advection $H_I$ as follows:
\begin{align}
U_I(t) &=e^{t H_I+t H_0}e^{-t H_0}. \label{def of UI}
\end{align}
$U_I(t)$ corresponds to the time-evolution operator in  the interaction representation based on the reference frame of $H_0$.
The time derivative of $U_I(t)$ is calculated as follows:
\begin{align}
\frac{\partial U_I(t)}{\partial t} &=e^{t H_I+t H_0}\biggl(H_I+H_0-H_0\biggr)e^{-t H_0},\\
&=e^{t H_I+t H_0}e^{-t H_0}\biggl(e^{tH_0}H_Ie^{-tH_0}\biggr),\\
&=U_I(t)H_I(t), \label{eq: EOM of UI}
\end{align}
where we have introduced $H_I(t)=e^{tH_0}H_Ie^{-tH_0}$. 
We note that the order of $U_I(t)$ and $H_I(t)$ on the r.h.s. is opposite to that in usual quantum mechanics.
The formal solution is given by
\begin{align}
U_I(t) =\tilde{\mathcal{T}}\exp\biggl[\int^t_0 d\tau H_I(\tau)\biggr],\label{eq:UIappe}
\end{align}
where $\tilde{\mathcal{T}}$ is the anti-time-ordered product:
\begin{align}
    \tilde{\mathcal{T}}\biggl[H_I(t_1)H_I(t_2)\biggr]=
\begin{cases}
H_I(t_1)H_I(t_2) & \mbox{ for } t_1 < t_2, \\
H_I(t_2)H_I(t_1) & \mbox{ for } t_1 > t_2. 
\end{cases}
\end{align}
$H_I(t)$ is calculated as 
\begin{align}
H_I(t)&=-\Gamma e^{tH_0}\bm{k}^2 e^{-tH_0},\\
&=-\Gamma \bm{q}(t)^2.\label{eq:HI}
\end{align}
Here, $\bm{q}(t) \equiv e^{tH_0} \bm{k} e^{-tH_0}$ is the advected wavenumber.
Because $H_I(t)$ is not the operator in the wavenumber space, we can drop $\tilde{\mathcal{T}}$ in  (\ref{eq:UIappe}).
Combining   (\ref{def of UI}), (\ref{eq:UIappe}), and (\ref{eq:HI}), we obtain
\begin{align}
  e^{t H_I+t H_0}=  \exp\biggl[-\Gamma\int^t_0 d\tau \bm{q}(\tau)^2\biggr]e^{t H_0}.
\end{align}
This is (\ref{eq:interaction picture}).


\section{Detailed calculation of \texorpdfstring{$\bm{q}(t)$}{TEXT}}
\label{sec:q(t)}
From the definition  (\ref{eq:def of q}), $\bm{q}(t)$  is written as 
\begin{align}
    \bm{q}(t)&=\mathcal{M}(t)\bm{k},\label{eq:qM}\\
    \mathcal{M}(t)&=\exp\biggl[\frac{t}{2}\mathcal{D}^T\biggr],
\end{align}
where the superscript $^T$ denotes the transpose of a matrix.
The time dependence of $\bm{q}(t)$ is determined by the matrix $\mathcal{M}(t)$.

To calculate $\mathcal{M}(t)$, we consider the eigenvalue $\lambda_\pm$ and the eigenvector $\bm{e}_\pm$ of $\D^T$, which are respectively given by
\begin{align}
\lambda_{\pm}&= \pm \sqrt{S^2-A^2}, \\
\bm{e}_{\pm}&=\frac{1}{\sqrt{2S}}
\begin{pmatrix}
\sqrt{S-A} \\
\pm\sqrt{S+A}
\end{pmatrix}.
\end{align}
The eigenvalue $\lambda_\pm$ is real for the elongational flow ($S>A$) and purely imaginary for the rotational flow ($A>S$).
For the shear flow, the eigenvalue is zero and degenerate.
From the eigenvectors, we obtain the diagonalizing matrix for $S \neq A$, 
\begin{align}
P&=\frac{1}{\sqrt{2S}}
\begin{pmatrix}
\sqrt{S-A} & \sqrt{S-A}\\
\sqrt{S+A} & -\sqrt{S+A}
\end{pmatrix}. \label{def P}
\end{align}
We then calculate $\mathcal{M}(t)$ as follows:
\begin{align}
\exp\biggl(\frac{t}{2}\D^T \biggr) &= P P^{-1} \exp\biggl(\frac{t}{2}\D^T \biggr) P P^{-1}, \\
&=P
\begin{pmatrix}
e^{t\lambda_+/2} & 0\\
0 & e^{t\lambda_-/2}
\end{pmatrix}
 P^{-1}, \\
&=
\begin{pmatrix}
\cosh (t \lambda_+/2) & \frac{S-A}{\lambda_+}\sinh (t\lambda_+/2)  \\
 \frac{S+A}{\lambda_+}\sinh (t \lambda_+/2) & \cosh (t \lambda_+/2),
\end{pmatrix},
\end{align}
where we have used $\lambda_+=-\lambda_-$.
Therefore, for the elongational flow $S>A$,  we obtain
\begin{align}
\mathcal{M}_e(t) =
\begin{pmatrix}
\cosh (t \Omega_e/2) & \frac{S-A}{\Omega_e}\sinh (t \Omega_e/2)  \\
 \frac{S+A}{\Omega_e}\sinh (t \Omega_e/2) & \cosh (t \Omega_e/2)
\end{pmatrix},
\label{eq:Me}
\end{align}
where we have introduced $\Omega_e=\lambda_+$. 
For the rotational flow $A>S$, $\Omega_e$ becomes $i\Omega_r=i\sqrt{A^2-S^2}$ and  hyperbolic functions also become  trigonometric functions.
Then, we obtain
\begin{align}
\mathcal{M}_r(t) =
\begin{pmatrix}
\cos (t \Omega_r /2) & \frac{S-A}{\Omega_r}\sin (t \Omega_r /2)  \\
 \frac{S+A}{\Omega_r}\sin (t \Omega_r /2) & \cos (t \Omega_r /2)
\end{pmatrix}.
\label{eq:Mr}
\end{align}
For the shear flow $A=S$, $\Omega_e$ becomes zero and $\mathcal{M}_e$ become
\begin{align}
\mathcal{M}_s(t) =
\begin{pmatrix}
1 & 0 \\
t S & 1
\end{pmatrix}.
\label{eq:Ms}
\end{align}

Therefore, the square of $\bm{q}_i(t)=\mathcal{M}_i(t)\bm{k}$ is calculated as 
\begin{align}
    \bm{q}_i^2(t) &=\bm{k}^2+\bm{k}^T \cdot \mathcal{F}_i(t) \cdot \bm{k},\label{eq:q in Appe}\\
      \mathcal{F}_s(t)&=\begin{pmatrix}
        t^2S^2& t S \\
        tS & 0
    \end{pmatrix},\label{eq:Fs in Appe}\\
     \mathcal{F}_e(t)&=\begin{pmatrix}
        \frac{2A}{S-A}\sinh^2(t\Omega_e/2) & \frac{2S}{\Omega_e}\sinh(\Omega_e t/2)\cosh(\Omega_e t/2) \\
        \frac{2S}{\Omega_e}\sinh(\Omega_e t/2)\cosh(\Omega_e t/2) &-\frac{2A}{S+A}\sinh^2(t\Omega_e/2)
    \end{pmatrix},\label{eq:Fe in Appe}\\
    \mathcal{F}_r(t)&=\begin{pmatrix}
        \frac{2S}{A-S}\sin^2(t\Omega_r/2) & \frac{2S}{\Omega_r}\sin(\Omega_r t/2)\cos(\Omega_r t/2) \\
        \frac{2S}{\Omega_r}\sin(\Omega_r t/2)\cos(\Omega_r t/2) &-\frac{2S}{A+S}\sin^2(t\Omega_r/2)
    \end{pmatrix}.\label{eq:Fr in Appe}
\end{align}
Furthermore, in the long-time limit, we obtain
\begin{align}
    \bm{q}_s^2(t)&\sim t^2S^2k_x^2,\label{eq:qs in assymptotic}\\
    \bm{q}_e^2(t)&\sim \frac{1}{2}e^{t\Omega_e} 
    \begin{pmatrix}
        k_x & k_y
    \end{pmatrix}
    \begin{pmatrix}
        S/(S-A) & S/\Omega_e \\
        S/\Omega_e & S/(S+A)
    \end{pmatrix} 
        \begin{pmatrix}
        k_x \\ k_y
    \end{pmatrix} \\
    &= e^{t\Omega_e} 
    \begin{pmatrix}
        k_{\mathrm{out}} & k_{\mathrm{in}}
    \end{pmatrix}
    \begin{pmatrix}
    S^2/\Omega_e^2 & 0 \\
        0 & 0
    \end{pmatrix} 
     \begin{pmatrix}
        k_{\mathrm{out}} \\ k_{\mathrm{in}}
    \end{pmatrix}, \label{eq:qe in assymptotic}
\end{align}
where we have diagonalized the matrix part of $\bm{q}_e^2(t)$ in the last line. We also introduce $k_{\mathrm{out}}$ and $k_{\mathrm{in}}$ as
\begin{align}
        \begin{pmatrix}
        k_{\rm out} \\ k_{\rm in}
    \end{pmatrix}
    &=\frac{1}{\sqrt{2S}}
    \begin{pmatrix}
        \sqrt{S+A} & \sqrt{S-A}\\
        -\sqrt{S-A} & \sqrt{S+A}
    \end{pmatrix}
        \begin{pmatrix}
        k_x \\ k_y
    \end{pmatrix},\\
\end{align}
which are the wavenumbers in the outgoing and incoming directions of the elongational flow, respectively.
The long-time behavior of $\bm{q}_r^2(t)$ does not change because it oscillates in time.

Finally, we consider the time integral of $\bm{q}_i^2(t)$, which are calculated as 
\begin{align}
\int^t_0 d\tau \bm{q}_s^2(\tau)&\sim \frac{1}{3}t^3S^2k_x^2,\\
\int^t_0 d\tau \bm{q}_e^2(\tau)&\sim \frac{S^2}{\Omega_e}e^{t\Omega_e}k_{\rm out}^2,\\
\int^t_0 d\tau \bm{q}_r^2(\tau)&\sim t\bm{k}^2.
\end{align}
We note that the term with  $\mathcal{F}_r(t)$ does not contribute to the time integral of $\bm{q}_r^2(t)$ in the long-time limit. 

\if
\section{Derivation of  (\ref{eq:static NG mode: list})}
\label{sec:appeD}
We first derive $\Cpps$ and $\Cppr$ of  (\ref{eq:static NG mode: list}) from $U_I^s$ and $U_I^r$ of  (\ref{eq:UI in long time}), respectively.
By changing the time variables as
\begin{align}
    \tau_s&=k_x^{2/3} t, \\
    \tau_r &= k^2 t,
\end{align}
we can extract the wavenumber dependences of the time integrals of $U_I^s$ and $U_I^r$
\begin{align}
    \int^\infty_0 dt e^{-\Gamma S^2 k_x^2 t^3} &= \frac{1}{k_x^{2/3}} \int^\infty_0 d\tau_s e^{-\Gamma S^2 \tau_s^3},\\
     \int^\infty_0 dt e^{-\Gamma k2 t} &= \frac{1}{k^2} \int^\infty_0 d\tau_r e^{-\Gamma \tau_r},
\end{align}
respectively. We note that the integral parts of the r.h.s do not depend on the wavenumber.
Then, we obtain $\Cpps$ and $\Cppr$ of  (\ref{eq:UI in long time}).

We next consider the time integral of $U_I^e(t)$.
By changing the time variable as
\begin{align}
  \tau_e=\Gamma S^2 \Omega_e^{-1}k_{\rm out}^2 e^{t\Omega_e},
\end{align}
we have
\begin{align}
    \int^\infty_0 dt \exp\biggl[-\Gamma S^2 \Omega_e^{-1} k^2_{\rm out} e^{t\Omega_e}\biggr]
    &=\frac{1}{\Omega_e}\int^\infty_{\Gamma S^2 \Omega_e^{-1} k^2_{\rm out}}\frac{e^{-\tau}}{\tau},\\
    &=-\frac{1}{\Omega_e}\mbox{Ei}(-\Gamma S^2 \Omega_e^{-1} k^2_{\rm out}), \label{eq:Ei}
\end{align}
where Ei is the exponential integral function
\begin{align}
    Ei(x)=-\int^\infty_{-x}d\tau \frac{e^{-\tau}}{\tau}.
\end{align}
The exponential integral function is represented as~\cite{abramowitz1968handbook}
\begin{align}
\mbox{Ei}(x)=\log x+\gamma-\mbox{Ein}(-x),
\end{align}
where $\gamma$ is the Euler constant and Ein$(x)$ is the entire function. 
Because Ein$(x)$ is the regular function, the singularity of (\ref{eq:Ei}) in the limit $k_{\rm out} \to 0$ is given by
\begin{align}
      \int^\infty_0 dt U_I^e(t)&=\int^\infty_0 dt \exp\biggl[-\Gamma S^2 \Omega_e^{-1} k^2_{\rm out} e^{t\Omega_e}\biggr],\\
    &\sim -\frac{1}{\Omega_e}\log k_{\rm out}^2.
\end{align}
Therefore, we arrive at $\Cppe$ of  (\ref{eq:static NG mode: list}).
\fi

\section{Procedure for drawing phase diagram}
\label{appendix:phase diagram}
To draw the phase diagram, we calculate the transition point $r^*$ without using the asymptotic behaviors in $t \to \infty$.
From   (\ref{eq:transitionpoint}), (\ref{eq:picorrelation_2}), and (\ref{eq:qM}), we write the transition point $r^* =-g C_{\pi\pi}$ as
\begin{align}
r^* &= 
-\frac{gT}{4\pi^2}\int d\bm{k}\int^\infty_0 ds \exp\biggl[- \bm{k}^T \cdot  \Gamma  \int^s_0  d\tau  \mathcal{M}^T_i \mathcal{M}_i  \cdot \bm{k} \biggr].
\label{eq:CwithM}
\end{align}

We note that  apart from the IR divergence, $C_{\pi\pi}$ shows the UV divergence, which originates from $|\bm{k}| \to \infty$. 
The UV divergence exists even in three-dimensional equilibrium systems
and is irrelevant to our purpose.
To regulate it, we introduce $e^{-\epsilon \bm{k}^2}$ with the infinitesimally small parameter $\epsilon$ as 
\begin{align}
r^*
= -gT\int \frac{d\bm{k}}{(2\pi)^2}\int^\infty_0 ds \exp\biggl[- \bm{k}^T \cdot  \Gamma \biggl( \int^s_0  d\tau  \mathcal{M}^T_i \mathcal{M}_i+\epsilon \mathcal{I} \biggr) \cdot \bm{k} \biggr], 
\end{align}
where $\mathcal{I}$ is the $2 \times 2$ unit matrix.
By changing the order of the $\bm{k}$ and $s$ integrals, we perform the Gaussian integral for $\bm{k}$ and obtain
\begin{align}
 r^*
&=-\frac{gT}{4\pi\Gamma}  \int^\infty_0 dt F_i(t),  \label{Cpipi} \\
F_i(t) &\equiv \det \biggl(\int^t_0  d\tau  \mathcal{M}^T_i \mathcal{M}_i +\epsilon \mathcal{I}\biggr)^{-1/2}. \label{eq:def of F}
\end{align}

From (\ref{eq:Me})--(\ref{eq:Ms}), $F_i(t)$ are calculated as
\begin{align}
    F_r(t) =&\biggl[\frac{2S^2(\cos(\Omega_r t)-1)+\Omega_r^2 A^2t^2}{\Omega_r^4} 
    +2\epsilon\biggl(t+\frac{t S^2}{\Omega_r^2}-\frac{S^2 \sin(\Omega_r t)}{\Omega_r^3}
    \biggr)\biggr]^{-1/2},\\
F_e(t) =&\biggl[\frac{2S^2(\cosh(\Omega_e t)-1)-\Omega_e^2 A^2 s^2}{\Omega_e^4}+2\epsilon\biggl(t-\frac{t S^2}{\Omega_e^2}+\frac{S^2 \sinh(\Omega_e t)}{\Omega_e^3}
\biggr)\biggr]^{-1/2},\\
F_s(t) =&\biggl[\frac{1}{12}S^2t^4+s^2+\epsilon\biggl(2t+\frac{1}{3}S^2t^3\biggr) \biggr]^{-1/2}.
\end{align}
We obtain $r^*$ numerically by performing the $t$ integral of $F_i(t)$.
The resulting phase diagram is given in Fig.~\ref{fig:phasediagram}.
We have set the parameters as $g=\Gamma=T=1$ and $\epsilon=10^{-3}$.

We also argue the UV divergence, which occurs in $\epsilon \to 0$.
The leading term of $F_i(t)$ in the limit $t \to 0$ and the integral of that are calculated as
\begin{align}
F_i(t) &\propto \frac{1}{ \sqrt{2\epsilon t}}, \label{eq:limit_s0} \\
\int_0^{\Lambda_0} dt  F_i(t) &\propto  \sqrt{\frac{\Lambda_0}{2\epsilon}},
\end{align}
where $\Lambda_0$ is the upper limit for the asymptotic behavior of $F_i(t)$.
This behavior is common among the three flows.
The integral diverges in the limit $\epsilon \to 0$, and thus, it is the UV divergence that we have regulated.
Consequently, the UV divergence appears from the short-time behavior.

\section{Derivation of  critical exponents}
\label{sec:scaling}
We derive the scaling behavior $\phiR \sim \delta^{1/4}$ with $\delta \equiv S/A-1$ in  (\ref{eq:exponent1}).
From  (\ref{eq:condensate}), (\ref{eq:transitionpoint}), and (\ref{Cpipi}), we see that the dependence of $\phiR$ on $S$ and $A$ originates from the integral of $F_i(t)$.

The asymptotic behavior of $F_e(t)$ in $t \to \infty$ and its integral are respectively calculated as
\begin{align}
F_e(s)&=\frac{\Omega_e^2}{\sqrt{2}S}e^{-s\Omega_e/2},\\
\int^\infty_{\Lambda_1} ds F_e(s)&=\frac{\sqrt{2}\Omega_e}{S}e^{\Lambda_1\Omega_e/2},
\end{align}
where $\Lambda_1$ is the cutoff for the asymptotic behavior.
By expanding the integral in $\delta$, we obtain the following in the leading order:
\begin{align}
\int^\infty_{\Lambda_1} ds F_e(s)= \frac{\sqrt{2}}{S}\delta^{1/2}.
\end{align}
Therefore, we have $\phiR^2 \sim \delta^{1/2}$ and thus obtain the scaling behavior in  (\ref{eq:exponent1}).

We next derive the scaling behavior of  (\ref{eq:exponent2}). 
From   (\ref{eq:condensate}) and (\ref{eq:transitionpoint}),
we obtain
\begin{align}
    \phiR=\frac{1}{g^{1/2}}(r-r^*)^{1/2}.
\end{align}
This is  (\ref{eq:exponent2}). 

\section{Time-evolution operator and interaction representation}
\label{sec:U(t)2}
We consider the time-correlation function  $D_{\pi\pi}(t,\bm{k})$ defined by 
\begin{align}
D_{\pi\pi}(t,\bm{k}) :=\langle \pi^\beta(t,\bm{k})\pi^\beta(0,-\bm{k})\rangle/L^2. 
\end{align}
By multiplying  (\ref{EOMpi3}) by $\pi^\beta(0,-\bm{k})$ and taking the noise average, we obtain
\begin{align}
\biggl[\frac{\partial }{\partial t}-\frac{1}{2}\bm{k} \cdot \D\cdot \nabla_{\bm{k}}+\Gamma \bm{k}^2\biggr]D_{\pi\pi}(t,\bm{k})  &=0.
\end{align}

\section{Superdiffusion and geometry of streamlines}
\label{sec:superdiff and geometry}
We demonstrate that whether the superdiffusion occurs is determined solely by the geometry of the streamlines without explicit calculations.
The key distinction among the rotational, shear, and elongational flows lies in the geometry of the streamline; as shown in Figs.~\ref{fig:q(t)}(a)--(c), it forms a closed curve in the rotational flow and extends towards infinity in the shear and elongational flows.

As explained in section~\ref{sec:physical interpretation}, the dynamics of the NG mode is described by the time evolution operator $U_I(t)$, which is given by (\ref{eq:decay of pi: rewrite}).
Then, the difference in the geometries of the streamlines is reflected in the dynamics of the NG mode through the functional form of $\bm{q}(t)$.
Because the advected wavenumber $\bm{q}(t)$ is the solution of the advection equations (\ref{eq:advection equation:1}) and (\ref{eq:advection equation:2}), $\bm{q}^2(t)$ is expressed as the quadratic form of $\bm{k}$:
\begin{align}
\bm{q}^2(t) =\bm{k}^2+ \bm{k}^T\cdot \mathcal{F} (t) \cdot \bm{k}, \label{eq:decomposed q}
\end{align}
where the first term represents the equilibrium part and $\mathcal{F}(t)$ is a $2\times 2$ matrix determined by the streamline in the wavenumber space. The expression of $\mathcal{F}(t)$ determines the behavior of $U_I(t)$, as shown in (\ref{eq:UI in long time}).
The concrete expressions of $\mathcal{F}(t)$ are given in (\ref{eq:Fs in Appe})--(\ref{eq:Fr in Appe}).

We first discuss the cases of shear and elongational flows, where the streamline goes towards infinity as shown in Figs.~\ref{fig:q(t)}(b) and (c).
The outgoing streamlines mean $\bm{q}(t)^2 \to \infty$ at $t \to \infty$, which in turn implies $\mathcal{F}(t) \to \infty$ at $t \to \infty$.
Then, the second term of (\ref{eq:decomposed q}) is dominant and the asymptotic behavior is given by
\begin{align}
\bm{q}^2(t) \sim \mathcal{F}_{\mathrm{out}}(t) k_{\mathrm{out}}^2,
\label{eq:extension of q(t)^2}
\end{align}
where $ k_{\mathrm{out}}$ and $\mathcal{F}_{\mathrm{out}}(t)$ are the components of $\bm{k}$ and $\mathcal{F}(t)$ in the outgoing direction of the streamline, respectively.
See (\ref{eq:qs in assymptotic}) and (\ref{eq:qe in assymptotic}) for the explicit expression of $\bm{q}^2(t)$ in the long-time limit.
By substituting (\ref{eq:extension of q(t)^2}) into (\ref{eq:decay of pi: rewrite}), we obtain
\begin{align}
U_I(t) \sim e^{-\Gamma F(t) k_{\mathrm{out}}^2}
\label{eq:UI(t) outwards}
\end{align}
with 
\begin{align}
    F(t) :=\int_0^t ds \mathcal{F}_{\mathrm{out}}(s).
\end{align}
Because $\mathcal{F}_{\mathrm{out}}(t) \to \infty$ at $t \to \infty$ due to the general feature of the outgoing streamline, its integral diverges faster than $t$. That is, we obtain
\begin{align}
\lim_{t \to \infty} \frac{F(t)k_{\mathrm{out}}^2}{t k^2}\to \infty,
\label{eq:behavior of F(t)}
\end{align}
which indicates the superdiffusion.

We next consider the case of   rotational flow.
In this case, in contrast to the shear and elongational flows, the streamline forms a closed curve such as that in Fig.~\ref{fig:q(t)}(a).
Because $\bm{q}^2(t)$ and $\mathcal{F}(t)$ solely oscillate in time, the time integral of $\mathcal{F}(t)$ also oscillates.
As a result, the first term of (\ref{eq:decomposed q}) is dominant and we have
\begin{align}
    \int^t_0 d\tau \bm{q}^2(\tau) \sim t\bm{k}^2.
\end{align}
Thus, we obtain the normal diffusion
\begin{align}
U_I(t) \sim e^{-\Gamma \bm{k}^2 t}.
\end{align}

\end{appendix}
\bibliography{
ref,
ref_flocking,
ref_nonreciprocal,
ref_advection,
ref_noneqcrystal,
ref_multitemperature,
ref_Levyflight,
ref_colorednoise,
ref_noneqfluctuations,
ref_criticalphenomena
}

\end{document}